\documentclass[twocolumn,prb,showpacs,superscriptaddress]{revtex4}
\usepackage{amssymb}
\usepackage{amsmath}
\usepackage{graphicx}
\usepackage{subfigure}
\usepackage{natbib}
\usepackage{epsfig}
\usepackage{amsfonts}
\usepackage{mathrsfs}
\usepackage{ulem}
\usepackage{color}
\usepackage[toc,page,title,titletoc,header]{appendix}

\begin{document}

\title{Topological nature of magnetization plateaus in periodically modulated quantum spin chains}
\author{Haiping Hu}
\affiliation{Beijing National Laboratory for Condensed Matter
Physics, Institute of Physics, Chinese Academy of Sciences,
Beijing 100190, China}
\author{Chen Cheng}
\affiliation{Center for Interdisciplinary Studies $\&$ Key Laboratory for
Magnetism and Magnetic Materials of the MoE, Lanzhou University, Lanzhou 730000, China}
\author{Zhihao Xu}
\affiliation{Beijing National Laboratory for Condensed Matter
Physics, Institute of Physics, Chinese Academy of Sciences,
Beijing 100190, China}
\author{Hong-Gang Luo}
\affiliation{Center for Interdisciplinary Studies $\&$ Key Laboratory for
Magnetism and Magnetic Materials of the MoE, Lanzhou University, Lanzhou 730000, China}
\affiliation{Beijing Computational Science Research Center, Beijing 100084, China}
\author{Shu Chen}
\thanks{Corresponding author, schen@aphy.iphy.ac.cn}
\affiliation{Beijing National
Laboratory for Condensed Matter Physics, Institute of Physics,
Chinese Academy of Sciences, Beijing 100190, China}
\affiliation{Collaborative Innovation Center of Quantum Matter, Beijing, China}
\begin{abstract}
We unveil nontrivial topological properties of zero-temperature magnetization plateau states in periodically modulated quantum spin chains under a uniform magnetic field.  As positions of plateaus are uniquely determined by the modulation period of exchange couplings, we find that the topologically nontrivial plateau state can be characterized by a nonzero integer Chern number and has nontrivial edge excitations. Our study clarifies the topological origin of the well-known phenomena of quantized magnetization plateaus in one-dimensional quantum spin systems and relates the plateau state to the correlated topological insulator.
\end{abstract}

\maketitle

\section{Introduction}

Interacting systems with topologically nontrivial properties are becoming important issues in recent studies of topological insulators \cite{review}. As the framework of topological insulators is based on the single-particle band theory, searching topological phases in correlated systems and characterizing them efficiently remain challenging problems \cite{WangZhong,Gurarie,Kitaev,Senthil,Pollmann2,Wen2011}.
Among various studies, topologically nontrivial phases in one-dimensional (1D) interacting spin systems have attracted particular attention \cite{Wen2011,Pollmann2}, as the spin chain systems have been typical strongly correlated systems exhibiting rich physical phenomena. A well-known example is the Haldane phase in the spin-1 chain \cite{haldane,AKLT}, which is a symmetry protected topological state, whereas the spin-$1/2$ chain with gapless excitations is not topologically protected. Although 1D spin-$1/2$ systems are generally lack of topological nontrivial phases, recently it was found that the free fermion systems trapped in optical superlattices exhibit topological non-trivial properties \cite{lang,kraus}, which can be characterized by a topological Chern number defined in an extended two-dimensional (2D) parameter space as the periodical modulation parameter provides an additional dimension.

Motivated by recent theoretical progress, in this work we shall study the topological properties of the 1D spin chain in a uniform magnetic field with exchange strengths between neighboring spins being periodically modulated. An interesting question is whether topologically nontrivial states induced by external periodic modulations can survive in the strongly correlated spin system? When the magnetic field is applied to a quantum spin chain, plateaus in the
zero-temperature magnetization curve are expected to appear \cite{hida}. In a pioneering work, Oshikawa, Yamanaka and Affleck (OYA) \cite{oshikawa} predicted that the magnetization plateaus should emerge when $m_z$ fulfilled the quantized relation $q(S-m_z)=integer$, where $q$ is the period of the ground state and $S$ is spin. Despite of lack of exact proof, the OYA's criterion for the appearance of magnetization plateaus has been confirmed from various aspects \cite{hida,oshikawa,tonegawa,totsuka,Honecker,exp,curve}.
Although the phenomena of quantum magnetization plateaus in 1D quantum spin systems are extensively studied, so far, few works \cite {nonlocal,z2z2,tkng,khida} have been concerned with the possible topological nature of these plateaus, which is the main purpose of the present work. By studying periodically modulated Heisenberg spin chains, we find that magnetization plateaus induced by the modulation are associated with excitation gaps and can be well characterized by non-zero Chern numbers defined in a 2D parameter space. Topologically nontrivial properties of plateau states are also demonstrated by the presence of edge states under open boundary conditions (OBCs), which locate in the gap regime and vary continuously with the variation of the modulation phase parameter. We investigate both spin-${1}/{2}$ and spin-$1$ chains and find some different behaviors associated with the existence of Haldane's gap for integer spin chains. Our results clarify the nontrivial topology of magnetization plateaus in spin chain systems and construct connection between magnetization plateaus and correlated topological insulators.

\section{Model and magnetization plateaus}
We consider the spin-$1/2$ Heisenberg model with periodically modulated coupling in a magnetic field, which is described by the following Hamiltonian
\begin{eqnarray}
H=\sum_i J_i(S_i \cdot S_{i+1})-h S_i^z, \label{xxx}
\end{eqnarray}
with
\begin{equation}
J_i = J [ 1- \lambda \cos(2\pi\alpha i+\delta)],  \label{Ji}
\end{equation}
where we take $\alpha=1/q$ with $q$ an integer and $h$ denotes the strength of the magnetic field. The coupling strength $J_i$ is periodically modulated via cosine modulations of the strength $\lambda$ with the periodicity $q$ and the phase factor $\delta$. The special
case with $\lambda=0$ reduces the Hamiltonian to the homogenous Heisenberg model. In this work, we shall focus on the case with antiferromagnetic couplings, i.e., $J>0$ and $|\lambda|<1$. For convenience, $J=1$ is taken as the unit of energy.

We first calculate the magnetization curve of the periodically modulated spin-$1/2$ Heisenberg chain with $\alpha=1/3$ by using the exact diagonalization method. The magnetization per spin is defined as $m_z =S_z/L$ with $S_z=\sum_i^L S_i^z$ being the $z$ component of the total spin and $L$ denoting the lattice size. In Fig.1, we display the magnetization $m_z$ with respect to the magnetic field $h$ for the spin chain with 3-period modulation, $L=24$ and $\delta=0$. It is clear that obvious magnetic plateaus appear at $m_z=1/2$, ${1}/{6}$, $-{1}/{6}$, and $-1/2$, which conforms to the OYA's criterion. We note that other small steps are due to the finite size effect \cite{appendix}, which shall disappear as the lattice size increases. Except for the saturation plateaus $m_z= \pm 1/2$, the appearance of non-vanishing plateaus is due to the existence of finite gaps in the thermodynamic limit with the width of plateaus proportional to the gap size. While the position of plateaus is irrelevant to the modulation phase $\delta$, the gap size (the plateau width) for the state with $m_z=\pm 1/6$ changes cyclically with a very small modulation amplitude when the phase $\delta$ varies continuously from $0$ to $2 \pi$.

\section{topological nature of plateaus}
As the cyclical variation of $\delta$ produces a family of systems with quite similar magnetization curves, we may topologically classify the plateau state of the family of systems with periodic modulation by assigning it to a topological Chern number \cite{Thouless,TKNN,niuqian}, which is defined in the 2D parameter space of $(\theta,\delta)$. Here $\theta$ is introduced by applying the twist boundary condition \cite{Shastry,huse,twist} to the spin operator, i.e.,  $S^{+}_{j+L}=e^{i\theta}S^{+}_{j}$, where $j$ denotes an arbitrary site. Explicitly, for spin models, the twist boundary condition can also be introduced by the following transformation, $H(\theta)=g(\theta)H g^{\dag}(\theta)$ with $g(\theta)=\prod_{j=1}^{L}e^{i S_j^z j \theta/L}$. Such a transformation only affects the XY term of the Heisenberg spin exchange interaction. In the 2D parameter space of $(\delta,\theta)$, the Chern number of the spin chain ground state is defined as an integral invariant
\begin{eqnarray}
C=\frac{1}{2\pi}\int d\theta d\delta F(\theta,\delta), \label{Chernnumber}
\end{eqnarray}
where $F(\theta,\delta)$ is the Berry curvature \cite{TKNN,niuqian} given by
\begin{eqnarray}
F(\theta,\delta)=Im(\langle\frac{\partial\psi}{\partial\delta}|\frac{\partial\psi}{\partial\theta}\rangle-
\langle\frac{\partial\psi}{\partial\theta}|\frac{\partial\psi}{\partial\delta}\rangle) .
\end{eqnarray}
The Chern number is only well defined as the ground state is protected by a finite gap and can be numerically calculated by using the method for a discrete manifold \cite{chern}. As shown in the inset of Fig.1, while states corresponding to saturate plateaus are topologically trivial, the plateau state corresponding to $m_z = \pm{1}/{6}$ has nonzero Chern number $\mp 1$. Similarly, for the case with $\alpha=1/5$, topological magnetic plateaus are found to appear at $m_z = {3}/{10}$, ${1}/{10}$, $-{1}/{10}$ and $-{3}/{10}$, which also fulfils the OYA's criterion. Correspondingly, Chern numbers for these plateau states are $-1$, $-2$, $2$ and $1$, respectively.
\begin{figure}
\includegraphics[width=3.7in]{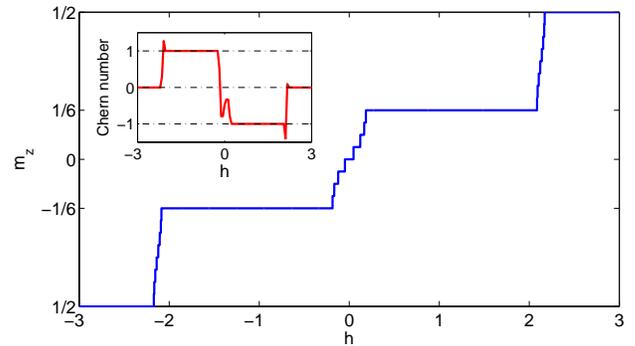}
\caption{(Color online) Magnetization curve of spin-$1/2$ Heisenberg chain for $\lambda =0.8$, $\alpha=1/3$, $\delta=0$ and $L=24$ under the periodic boundary condition (PBC). Inset shows the Chern number of the ground state versus magnetic field $h$.}
\end{figure}

It is worth mentioning that some previous works have used Berry phase \cite{voit,nakamura,hatsugaiberryphase,xiaohu}, which is an integral only over twist angle $\theta$, to characterize different insulating phases.
The relation between the Lieb-Schultz-Mattis (LSM) operator and Berry phase, which is crucial to OYA's argument to locate the plateaus, has been also discussed in Ref.\cite{voit,nakamura}, where the Berry phase is recognized as a phase angle of the twist operator appearing in LSM's argument \cite{LSM}. This phase can take two distinct values $0$ or $\pi$(mod $2\pi$) which cannot classify different plateau states in our present work obviously. The topological invariant we defined further attributes an integer number for each specific magnetization plateau which totally characterizes the topological nature for the corresponding plateau state. In the bosonization language, there is an analogy between Mott insulating states and the magnetization plateaus in spin chains. The above Chern number can be interpreted as numbers of spin-flip excitations pumped in one cycle of $\delta$ \cite{Thouless}, which is analogous to charge polarization \cite{resta} in Mott insulating phase.

For topologically nontrivial plateau states, we may expect to observe edge states under OBCs according to the bulk-edge correspondence. In practice, existence of nontrivial edge states is usually considered to be a hallmark of non-trivial topological properties even for a correlated topological state \cite{Wen,Zhu,WangYF}. To illustrate it clearly, we consider states with topological plateaus $m_z={3}/{10}$ and ${1}/{10}$ for the system with $\alpha=1/5$ and $L=90$ and calculate quasi-particle excitations around the plateau states by using the density matrix renormalization group (DMRG) method under OBCs. Since $S_z = N- L/2$ is a conserved quantity, where $N$ denotes the number of up spins. The plateau states with $m_z={3}/{10}$ and ${1}/{10}$  correspond to $N=72$ and $N=54$, respectively. Define $E_N$ as the ground state energy with $N$ up spins and $\Delta E_N= E_{N+1}-E_{N}$ as the excitation energy by flipping a spin from spin-down to spin-up. In Fig.2(a) and Fig.2(b), we show the excitation spectrum as a function of the phase $\delta$ for cases with $m_z={3}/{10}$ and $m_z={1}/{10}$, respectively. For both cases, it is clear that there is an obvious excitation gap when deviating from magnetization plateaus. Inside the gap regime, there exist two branches of excitation modes which cross each other and connect the lower and upper branch of excitation spectrums when $\delta$ varies from $0$ to $2\pi$.

In Fig.2(c) and Fig.2(d), we display the corresponding spin-flip distributions of the in-gap excitation modes by calculating $\Delta\rho_N=\rho_{N+1}-\rho_N$, where $\rho_N$ is defined as $\rho_N(i)=\langle\psi| S_i^z|\psi\rangle$ with $\psi$ being the ground state wave function with fixed $N$ up spins.
As illustrated in the figures, distributions of spin flipping excitations locate either on the left or right edge of the chain, which clearly indicates the in-gap excitation modes to be edge modes. Varying $\delta$ continuously, one can adiabatically transport the edge state from one side of the chain to the other side. While there is only one edge transport channel in the gap regime for the case of $m_z=3/10$ in Fig2.(a), there are two edge transport channels for the case of $m_z=1/10$ in Fig.2(b), which also reveals the topological feature of bulk states characterized by different Chern numbers. We have also analyzed edge modes and spin distributions carefully for systems with different sizes and do not find obvious differences \cite{appendix}.
\begin{figure}
\includegraphics[width=3.7in]{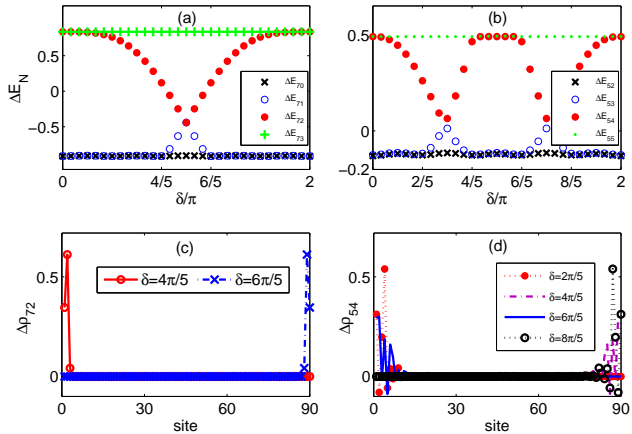}
\caption{(Color online) Excitation spectrum for plateau states with $m_z=3/10$ (a) and $m_z=1/10$ (b) under OBCs. (c) and (d) denote the corresponding spin-flip density distribution for the in-gap edge modes with various $\delta$. We take $\lambda=0.8$, $L=90$, $h=1.8$ ((a), (c)) and $h=0.3$ ((b), (d)) in our DMRG calculation.}
\end{figure}

Next we study the effect of anisotropic exchange coupling on the topological plateau states. To this end, we consider the periodically modulated XXZ model described by
\begin{eqnarray}
H=\sum_i J_i(S_i^x S^x_{i+1}+S_i^y S^y_{i+1}+ \gamma S_i^z S_{i+1}^z)-h S_i^z, \label{xxz}
\end{eqnarray}
where $J_i$ is given by Eq.(\ref{Ji}) and $\gamma$ describes the anisotropy of spin exchange coupling. When $\gamma=1$, the Hamiltonian reduces to the isotropic Heisenberg model of (\ref{xxx}). By using the DMRG method, we calculate the above XXZ model with $L=90$ under periodic boundary conditions and display magnetization curves in Fig.3(a) and Fig.3(b) for systems with $\alpha=1/3$  and $\alpha=1/5$, respectively. As shown in the figures, the magnetization curves for systems with different $\gamma$ ($\gamma=0.5$, $1$, and $2$) have similar structures. While the plateau heights for different $\gamma$ are identical, we observe that the plateau width becomes wider as $\gamma$ increases. The magnetization plateaus do not vanish even in the limit of $\gamma=0$. The presence of a finite width (energy gap) in the whole regime of $\gamma>0$ suggests that the plateau states with the same $m_z$ but different $\gamma$ have the same topological properties characterized by the same Chern number.
\begin{figure}
\includegraphics[width=3.7in]{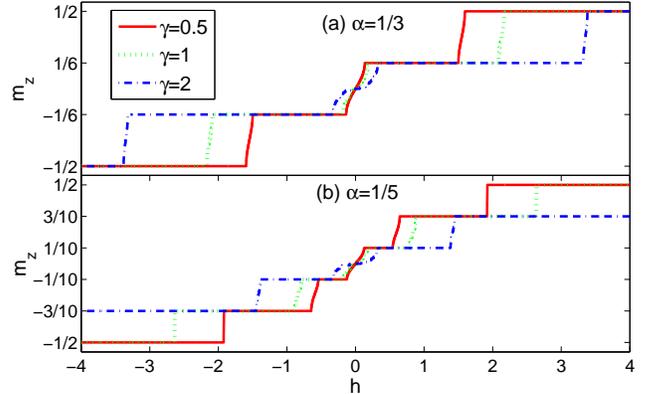}
\caption{(Color online) Magnetization curves of the spin-$1/2$ XXZ chain for different $\gamma$ with  $\lambda=0.8$, $\delta=0$ and $L=90$ under PBCs. (a) $\alpha=\frac{1}{3}$; (b) $\alpha=\frac{1}{5}$.}
\end{figure}

Particularly, the model in the limit of $\gamma=0$ (modulated XX model) can be mapped to a free fermion model with periodically modulated hopping amplitudes $J_i/2$ after a Jordan-Wigner transformation. For a general case with $\alpha=p/q$ (p and q are co-prime integers), the energy spectrum splits into $q$ bands. In Fig.4(a) and Fig.4(b), we display the spectrum versus $\alpha$ for the system with phase $\delta=0$ and $\delta=\pi/2$, respectively. The family of Hamiltonian
\[
H(\delta)= \sum_i J_i(\delta) (S_i^x S^x_{i+1}+S_i^y S^y_{i+1})
\]
has similar butterfly structure of spectrum, which can be mapped to the Hofstadter spectrum if $\delta$ is viewed as $k_y$ \cite{Hofstadter,Ganeshan,Kohmoto}. Tuning magnetic field is equivalent to sweeping the chemical potential in the language of free fermions, and magnetization plateau emerges when the chemical potential lies in the gap. The number of plateaus directly reflects the number of bands of our system in the limit of $\gamma=0$. We observe that the obtained Chern numbers have exact correspondence with the heights of magnetization plateaus: the plateaus can only appear at some specific heights which are completely determined by $\alpha$, i.e., $m_z=-1/2+(\alpha,1-\alpha,2\alpha,1-2\alpha...)$ if the values of $m_z$ are in the range of $(-1/2,1/2)$. The Chern number can be directly obtained by the Streda formula \cite{Streda}
\[
C= \frac{\partial m_z}{\partial\alpha},
\]
which gives $C=1,-1$ for $m_z=-1/2+(\alpha,1-\alpha)$ and $C=2,-2$ for $m_z=-1/2+(2\alpha,1-2\alpha)$.

The nontrivial edge excitations under the OBC can be also understood based on the band structure of the system in the limit of $\gamma=0$. As an example, we show the single particle spectrum with respect to $\delta$ in Fig.4(c) for the system with $\alpha=1/5$ and $L=90$ under the OBC. The plateau state with $m_z = 3/10$ corresponds to the state with four bands being filled, and thus it is straightforward to get the excitation spectrum around the plateau state as displayed in Fig.4(d). In comparison with Fig.2(a), it is not strange to find that they have similar structure as these plateau states can be adiabatically connected by varying $\gamma$.
\begin{figure}
\includegraphics[width=3.5in]{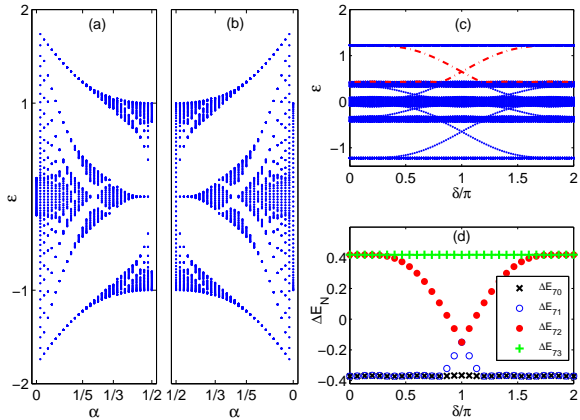}
\caption{(Color online) Energy spectrum with respect to $\alpha$ for the spin-1/2 XX model with $\lambda=0.8$, $L=90$ and different modulation phases: (a) $\delta=0$ and (b)$\delta=\pi/2$. (c) Single-particle spectrum for $\alpha=1/5$ with respect to phase $\delta$ under the OBC. (d) Excitation spectrum for the plateau state of $m_z=3/10$.}
\end{figure}

\section{Spin-$1$ model with periodic modulation}
Finally, we consider the periodically modulated spin-$1$ Heisenberg chain with $J_i$ given by Eq.(\ref{Ji}) and demonstrate that the plateau induced by the periodic modulation is topologically nontrivial in the same sense as the spin-1/2 model.
As shown in Fig.5(a) for the system with $\alpha=1/3$, $\lambda=0.3$ and $L=90$, the magnetization plateaus appear at $m_z=\pm1$, $\pm {2}/{3}$, $ \pm {1}/{3}$, and $0$, which is consistent with OYA's criterion. To reveal the nontrivial topological properties, we display the excitation spectrum for the plateau states with $m_z=0$, ${1}/{3}$, and ${2}/{3}$ under the OBC as a function of the phase $\delta$ in Fig.5(b), (c), and (d), respectively. Here we define $\Delta E_{S_z}=E_{S_z+1}-E_{S_z}$ and $\Delta\rho_{S_z}=\rho_{S_z+1}-\rho_{S_z}$, where $\rho_{S_z}(i)=\langle\psi| S_i^z|\psi\rangle$ with $\psi$ and $E_{S_z}$ the ground state wave function and energy in the $S_z$ subspace. For plateaus with $m_z=1/3$ and $2/3$, there exist continuous edge states which connect the the lower and upper branches of excitation bands.
However, for the plateau state with $m_z=0$, the edge excitations exhibit quite different behaviors from cases of $m_z={1}/{3}$ and ${2}/{3}$ as edge modes in the gap regime do not connect the lower and upper branches of excitation bands. Actually, the plateau with $m_z=0$ is not induced by the modulation but related to the Haldane gap for integer spin chains, and thus it has zero Chern number according to the definition of Eq.(\ref{Chernnumber}).
\begin{figure}
\includegraphics[width=3.6in]{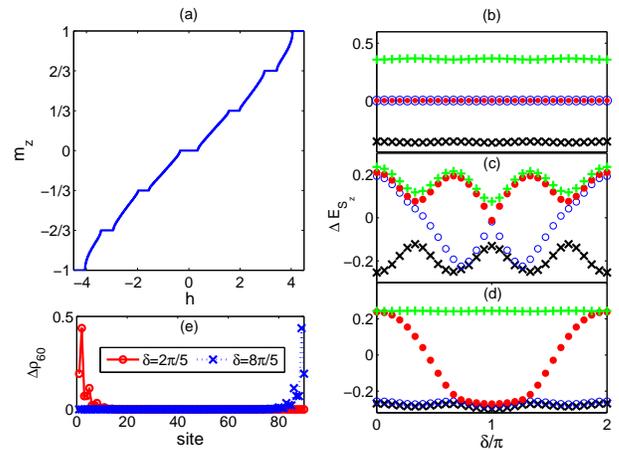}
\caption{(Color online) (a) Magnetism plateaus for the modulated spin-1 Heisenberg chain with $\lambda=0.3$ and $L=90$. (b),(c),(d) Excitation spectrum for plateau states with various $m_z$ under OBCs, i.e., $m_z=0$ ($h=0$), $m_z=1/3$ ($h=1.8$), and $m_z=2/3$ ($h=3.2$), respectively. Symbols of `plus', `filled dot', `open dot' and `cross' denote $\Delta E_{S_z}$ with $S_z = m_p L +1$, $ m_p L$, $m_p L-1$, and $m_p L-2$, respectively. Here $m_p$ represents the plateau height. (e) Spin-flip density distributions of edge modes for plateau states with $m_z=2/3$.}
\end{figure}

Plateau states in the modulated spin-1 chain can be also classified by Chern numbers. For the above case of $\alpha=1/3$, from bottom to top plateau, Chern numbers are $0,1,2,0,-2,-1,0$. Our results reveal the following rules for Chern numbers of available plateau states, i.e., $C=0$ for $m_z=0,\pm 1$ plateaus and $C=(1,-1,2,-2...)$ for $m_z=-1+(\alpha,2-\alpha,2\alpha,2-2\alpha...)$. We can clearly see, the plateaus $m_p$ and counting rules of Chern number for both spin-1/2 and spin-1 cases can be unified by the following formula: $m_p= \pm (n \alpha -S)$ with  $n=1,2, \cdots$
and $C_{m_p}=-sgn(m_p)q(S-|m_p|)$, where $sgn(m_p)$ is the sign function.
For the topologically nontrivial plateaus, we find that the generalized Streda formula $C_{m_p}= \frac{\partial m_p}{\partial\alpha}$ remains valid. The modulated higher-spin chains, e.g., $S={3}/{2}$, should demonstrate similar behaviors and we believe that the modulation induced topological plateaus in higher-spin systems can be described in similar schemes.

\section{Summary} In summary, we have studied the magnetization plateau states in quantum spin chains with periodically modulated couplings and revealed their nontrivial topological properties. The topological plateau states can be characterized by non-zero Chern numbers defined in the extended 2D parameter space and exhibit nontrivial edge excitations under the OBC. Particularly, for the modulated anisotropic spin-1/2 XXZ chain, we show that the topological plateau state protected by a finite excitation gap can be adiabatically connected to the plateau state in the XX limit, whose topological properties can be well understood based on the band theory. Our study gives a straightforward interpretation for the topological origin of quantized magnetization plateaus in 1D modulated quantum spin systems.

\begin{acknowledgments}
This work is supported by National Program for Basic Research
of MOST(973 grant), NSFC under Grants No.11374354, No.11174360, No.11174115, No.10974234 and No.11325417, and PCSIRT (Grant No.IRT1251).
\end{acknowledgments}

\appendix
\section{Comparison between DMRG and ED results}
To show the validity of the density matrix renormalization group (DMRG) method on the calculation of the edge excitations of our model, in this appendix we provide numerical results by using the exact diagonalization (ED) method for the system with the lattice size of $L=24$, and make a comparison between the ED results and the numerical results obtained by the DMRG method for the same system.

In Fig.6(a), we show the excitation spectrum for the plateau state with $m_z=1/6$ of the periodically modulated spin system with $\alpha=1/3$ and $L=24$ under the open boundary condition (OBC). The corresponding spin distributions obtained by the ED method is shown in Fig.6(c). The numerical ED results clearly indicate that the edge excitation modes change continuously and connect the lower and upper branch of excitation spectrums when $\delta$ varies from $0$ to $2\pi$. As the plateau state with $m_z=1/6$ is characterized by the Chern number $C=-1$, it is not strange that its edge excitation spectrum and the corresponding spin distributions have similar structures as the plateau state with $m_z=3/10$ of the periodically modulated spin system with $\alpha=1/5$ shown in Fig.2(a) and Fig.2(c) of the main text, which is also characterized by the Chern number $C=-1$. The ED results show that the topological features have already been definitely observed even for the system with lattice size of $L=24$. In Fig.6(b) and Fig.6(d), we also show the excitation spectrum and the corresponding spin distributions for the same plateau state calculated by the DMRG method. As expected, our DMRG results are almost identical to the ED results shown in Fig.6(a) and Fig6(c). Such a comparison clearly indicates that our DMRG calculation (including boundary quantities) is rather reliable. We also note that the difference of ground state energies calculated from the ED and DMRG methods is less than $10^{-8}$.
\begin{figure}[tbp]
\includegraphics[width=3.6in] {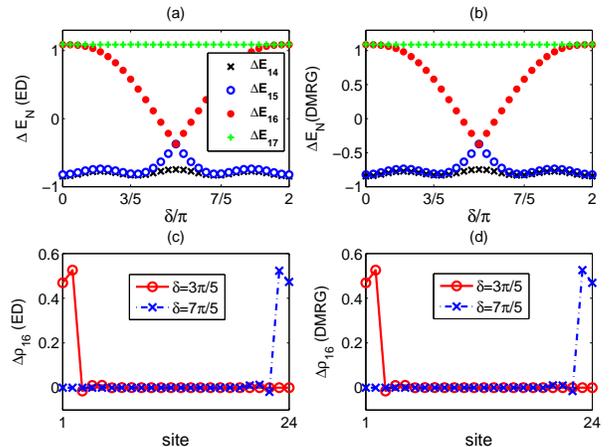}
\caption{(Color online) Excitation spectrum versus $\delta$ for the plateau state with $m_z=1/6$ ((a) and (b)) for the system with $L=24$ under the OBC. (c) and (d) denote the corresponding spin-flip density distributions for the in-gap edge modes with various $\delta$. (a) and (c) ((b) and (d)) are calculated via the ED (DMRG) method. Here $\lambda=0.8$, $h=1$ and $\gamma=1$.}
\end{figure}

\section{Finite size effect}
In the main text, we have mentioned the small step in the magnetization curve shown in Fig.1 of the main text is due to the finite size effect while the topological plateaus are quite robust and do not vanish with the increase of the lattice size. In this supplementary material, we show the dependence of plateau width on the length of spin chains and carry out the finite size analysis.
 We consider the case with $\alpha=1/3$ and take the topologically non-trivial plateau state at $m_z=1/6$ and the trivial plateau state at $m_z=0$ as examples. For a specific plateau, we denote $h_{low}$ and $h_{up}$ as the lower and upper transition point in the magnetization curve, respectively, and then $h_{width}=h_{up}-h_{down}$ is the width of the plateau. We calculate systems with different lattice sizes and illustrate the main results in Fig.7. As shown in Fig.7(a), the width of the plateau state with $m_z=1/6$ tends to a finite value as $L \rightarrow \infty$. On the contrary, the plateau width for the state with $m_z=0$ tends to zero in the limit of  $L \rightarrow \infty$ as shown in Fig.7(b).
 The non-vanishing plateau due to a finite gap in the thermodynamic limit is necessary for the robustness of topological properties of the plateau state.
\begin{figure}[tbp]
\includegraphics[width=3.6in] {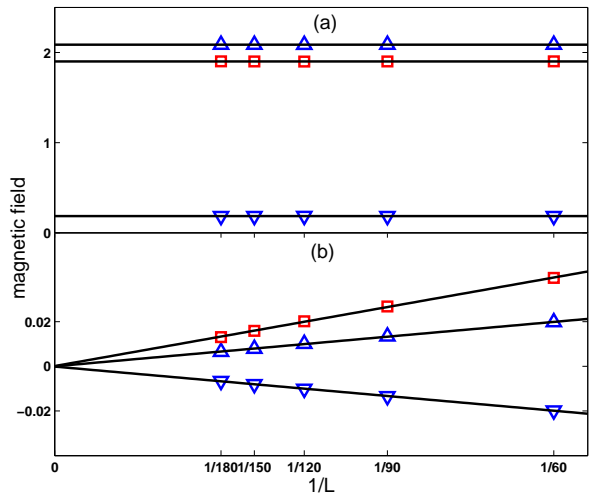}
\caption{(Color online) Dependence of the plateau width on the lattice size $L$. (a) is for the topological plateau with $m_z=1/6$, and (b) is for the small trivial plateau at $m_z=0$. Blue upper (lower) triangles represent upper (lower) transition points, red rectangles represent the plateau widthes. The black lines come from the linear fitting. Here we have taken $\lambda=0.8$, $\alpha=1/3$, $\delta=0$ and $\gamma=1$ under the periodic boundary condition. }
\end{figure}

\end{document}